\documentclass[a4paper,twoside]{article}

\usepackage{graphicx}
\usepackage{subcaption}
\usepackage{calc}
\usepackage{amssymb}
\usepackage{amstext}
\usepackage{amsmath}
\usepackage{booktabs}
\usepackage{multirow}
\usepackage{array}
\usepackage{natbib}
\usepackage{algorithm2e} 
\usepackage[bottom]{footmisc}
\usepackage{SCITEPRESS}

\usepackage[T1]{fontenc}

\newcommand{\dcintroimage}[1]{\includegraphics[height=0.85cm]{figures/#1}}
\newcommand{\dcimage}[1]{\includegraphics[height=0.65cm]{figures/#1}}
\newcommand{\dcevalimage}[1]{\includegraphics[height=0.85cm]{figures/#1}}
\newcommand{\heatmap}[1]{\includegraphics[width=0.145\textwidth]{#1}}
\def\AgentSet{\mathcal{I}}
\def\Actions{\mathcal{A}}
\def\AgentCountTrain{N_{\mathrm{train}}}
\def\AgentCountEval{N_{\mathrm{eval}}}
\def\DCSet{\mathcal{D}}
\def\PosInt{\mathbb{Z}_+}

\begin{document}

\title{Post-Training Policy Replication for Strategy-Following Multi-Agent Deep Reinforcement Learning across Team Sizes}

\author{\authorname{Yamato Takahagi\sup{1}, Yoshinari Motokawa\sup{1} and Toshiharu Sugawara\sup{1}}
\affiliation{\sup{1}Department of Computer Science and Communications Engineering, Waseda University, 3-4-1 Okubo, Shinjuku-ku, Tokyo 169-8555, Japan}
\email{\{y.takahagi, y.motokawa\}@isl.cs.waseda.ac.jp, sugawara@waseda.jp}
}

\keywords{Multi-Agent Deep Reinforcement Learning, Strategy Following, Policy Replication, Variable Team Size, Destination Channel.}

\abstract{
Although multi-agent deep reinforcement learning enables coordinated behavior, aligning it with managerial intent remains difficult due to complex agent interactions. Although the \emph{strategy-following distributed attentional actor architecture} (sfDA7-X) improves controllability via regional instructions, instruction combinations scale exponentially with team and instruction sizes, leading to prohibitive training costs and poor transferability. To reduce this computational cost, we introduce a team-size-independent observation architecture and reconstruct collective behavior by replicating a single trained policy across agents in teams of different sizes, thereby eliminating the need to train from scratch. Additionally, we evaluate warm-start retraining by initializing policy networks with a pre-trained policy to significantly lower computational overhead. In an \emph{object collection game}, a policy trained in a four-agent team is deployed to teams of three to eight agents. Experiments demonstrate that training with the proposed aggregated observation matrix achieves performance comparable to that of the baseline method, sfDA7-X. Furthermore, policy replication preserves performance without substantial degradation in three- and five-agent teams, and warm-start retraining outperforms training from scratch while reducing required training episodes by up to $90\%$. Deployments to six- and eight-agent teams further confirm effective spatial region allocation following given instructions. These results demonstrate that the proposed method enables zero-shot collective behavior reconstruction via policy replication.}

\onecolumn \maketitle \normalsize \setcounter{footnote}{0} \vfill

\section{\uppercase{Introduction}}\label{sec:translation-introduction}
In recent years, multi-agent systems, in which multiple autonomous agents collaborate to execute complex tasks, have attracted substantial research interest. Multi-agent deep reinforcement learning (MADRL) offers a powerful framework for acquiring coordinated behaviors directly from environmental observations, showing strong potential in applications such as autonomous driving \cite{shalev2016safe}, drone network management~\cite{shi2024task}, and strategy games~\cite{samvelyan2019starcraft}. However, intricate coordination patterns learned by MADRL make it challenging for human managers to align collective behaviors with dynamic operational requirements. Particularly in spatial tasks, such as patrolling and collection---where agents divide a shared workspace---system controls must enable flexible behavior adjustments without creating coverage gaps while maintaining robust coordination under reassigned work regions or varying team sizes.
\par

Several studies have aimed to improve the controllability of reinforcement learning agents through human guidance, natural language instructions, or program-guided policies \cite{wu2023toward,fried2018speaker,sun2020program}. However, these approaches primarily focus on single-agent settings. In a multi-agent setting, agents must not only follow their respective instructions but also coordinate to avoid operational overlaps and prevent unassigned coverage gaps. Consequently, single-agent methods cannot be directly applied.
\par

To extend controllability to multi-agent systems, \citet{motokawa2023strategy} proposed the {\em strategy-following distributed attentional actor architecture after conditional attention} (sfDA6-X). This framework incorporates a {\em destination channel} (DC)---a matrix representing an agent's designated work region---to control agent behavior without specifying explicit action sequences, where X denotes the type of attention head. By modifying the DC, a manager can provide trained agents with high-level regional instructions, thereby introducing controllability into MADRL. Building upon this, \citet{takahagi2025strategy} proposed the sfDA7-X architecture, which complements the standard DC with an Aggregated DC (ADC), which aggregates the DCs assigned to all other agents. By observing both its own DC and the ADC, sfDA7-X enables agents to accommodate biased instruction distributions and spontaneously cover unassigned regions.
\par

However, this DC-based framework suffers from combinatorial explosion. As the number of candidate DC types $K$ and agents $N$ increase, the number of possible instruction assignments grows exponentially up to $K^N$. Consequently, exposing agents to such diverse instruction configurations requires prohibitive training time, making frequent retraining under changing team sizes or operational requirements infeasible for practical applications. Furthermore, because sfDA7-X trains distinct agent-specific policies, it lacks a mechanism for directly deploying trained policies to teams of different sizes. Although recent studies have investigated zero-shot generalization to dynamic team compositions by incorporating robustness to team size changes into training \cite{ye2023mutual,guo2025heterogeneous}, instruction-based controllability remains beyond their scope.
\par

This study extended sfDA7-X by introducing an agent-count-independent observation representation. By leveraging this representation, we can select a single trained policy and replicate it across all agents, enabling zero-shot deployment to teams of varying sizes without additional training overhead. This formulation eliminates observation-shape incompatibilities caused by changes in team size, regardless of agent identities. Consequently, we can evaluate whether coordinated behavior emerges at team sizes different from those in the training configuration, bypassing the need for computationally expensive retraining while preserving the agents' capability to interpret managerial instructions. 
\par

Using an {\em object collection game}, we first compared the four-agent learning curves of sfDA7-X and the proposed architecture with aggregated observations using the same DRL head before policy replication. We then evaluated policy replication by deploying a four-agent policy to three- to five-agent teams and comparing the performance with that of target-size baselines trained individually under the same architecture. Averaged over the source policies, the three-agent configuration achieved collection rates comparable to those of the baseline, with a higher overall reward. Although the performance decreased slightly for four and five agents, no severe degradation was observed. Furthermore, selecting optimal source policies allowed three- to five-agent teams to outperform the baseline method. Beyond zero-shot replication, we evaluated warm-start retraining for three and five agents by initializing training with the best-performing four-agent source policy. Finally, deploying four-agent policies to six- and eight-agent teams, where training new baselines is computationally demanding, qualitatively confirmed instruction-conditioned region specialization via collection-location heatmaps. Overall, this study demonstrates both zero-shot collective behavior reconstruction and the efficacy of target-size warm-start retraining through single-policy replication.
\par

\section{\uppercase{Related Work}}\label{sec:translation-previous}
Prior studies have controlled a reinforcement learning agent through human guidance, natural language instructions, and program-guided policies~\cite{wu2023toward,fried2018speaker,sun2020program}; however, these studies primarily address single-agent environments. In contrast, strategy-following MADRL, such as sfDA7-X~\cite{takahagi2025strategy}, aims to coordinate multiple agents so that spatial instructions do not produce redundant overlapping work or uncovered regions unless explicitly instructed by a manager. To address this issue, sfDA7-X provides each agent with its regional instruction and the aggregated instructions assigned to the other agents during training and execution.
\par

Several recent studies have proposed learning methods that adapt to changes in the team size. For example, \citet{ye2023mutual} investigated zero-shot generalization for dynamic team compositions by introducing mutual information as an augmented reward, thereby preventing individual policies from over-relying on team-specific information. They derived a multi-agent policy iteration framework with a fixed marginal distribution and leveraged an imaginary team composition distribution for practical optimization. Their experiments demonstrated enhanced generalization to unseen team compositions in cooperative tasks.
\par

In addition, \citet{guo2025heterogeneous} proposed a scalable and heterogeneous \emph{proximal policy optimization} variant that introduces agent-specific strategy patterns into parameter-shared MARL via latent representations, enabling zero-shot scaling in the {\em StarCraft multi-agent challenge} (SMAC)~\cite{samvelyan2019starcraft} and Google Research Football~\cite{kurach2020google}. Similarly, \citet{wang2024nahteamwork} formulated $N$-agent ad hoc teamwork, where controlled agents cooperate with dynamically changing teammates by learning teammate behavioral representations for policy optimization. Experiments in the multi-agent particle environment and SMAC II~\cite{SMAC2} demonstrated effective generalization to unseen team members.
\par

Although these studies enhance the robustness of team composition, scalable collaboration, and adaptation to unknown teammates, their formulations do not provide execution-time controllability, where a manager or user supplies explicit instructions to intentionally guide or alter agent behavior post-training. Because such instruction-conditioned control is a core requirement of the present study, these methods cannot be directly adopted as substitutes for strategy-following MADRL. However, the original sfDA7-X framework relies on agent-count-dependent input structures, preventing direct policy transfer to teams of different sizes. Therefore, we extend the sfDA7-X framework by introducing an agent-count-independent observation architecture, enabling zero-shot deployment and warm-start retraining across varying team sizes while fully preserving its spatial instruction-following capabilities.
\par

\section{\uppercase{Preliminary}}\label{sec:translation-preliminaries}
\subsection{Object Collection Game}
We consider a discrete-time formulation ($t \in \{0, 1, 2, \dots\}$). Our method is evaluated using the {\em object collection game}, which is formulated as a {\em decentralized partially observable Markov decision process} (Dec-POMDP) defined for a set of $N$ agents $\AgentSet=\{0,\ldots,N-1\}$~\cite{bernstein2002complexity}. The Dec-POMDP is specified by the set of all possible states $\mathcal{S}$, action space $\Actions_i$ for each agent $i\in\AgentSet$, state transition probability $p_T$, reward function $r_i$, observation space $\Omega_i$, observation probability $O$, and episode horizon $H$. At time step $t$, the joint action of all agents is denoted by $a_t\in\prod_{i\in\AgentSet}\Actions_i$, and the environmental state is $s_t\in\mathcal{S}$. Agent $i$ selects an action according to its policy $\pi_i$ based on its local observation $o_{i,t}\in\Omega_i$, aiming to maximize the expected discounted return
\begin{equation}
  R_i=\sum_{t=0}^{H}\gamma^t r_i(s_t,a_t),
  \label{eq:translation-return}
\end{equation}
where $\gamma\in[0,1)$ is the discount factor.
\par

The object collection game is played on a $G_X\times G_Y$ grid ($G_X, G_Y\in\PosInt$, where $\PosInt$ denotes the set of positive integers). A fixed number of objects are placed at the grid cells. Each agent $i\in\AgentSet$ starts from a predefined initial position and executes an action $a_{i,t} \in \Actions_i=\{\mathrm{up},\mathrm{down},\mathrm{left},\mathrm{right}\}$ at every time step. We denote the node occupied by agent $i$ at time $t$ by $c_i(t)$. An agent that enters a node containing an object collects it and receives a positive reward $r_o$. Upon collection, the object vanishes, and a new object is spawned at another node. A collision with a wall or another agent yields a negative reward $r_c$. Specifically, a collision occurs at time $t$ if agent $i$ attempts to move into a wall, if two distinct agents occupy the same node ($c_i(t)=c_j(t)$ for $i \neq j$), or two agents swap positions, that is, $(c_i(t-1)=c_j(t))\wedge(c_j(t-1)=c_i(t))$. Any other movement incurs a small movement cost $r_m\leq 0$. The objective is to collect as many objects as possible at minimal cost while avoiding collisions.
\par

\subsection{Local Observation}
Agent $i$ constructs {\em observation matrices} from an $R_X\times R_Y$ local region centered on its current position, where $R_X$ and $R_Y$ are odd integers satisfying $3\leq R_X\leq G_X$ and $3\leq R_Y\leq G_Y$. The original sfDA7-X generates $N_c=N+2$ matrices, which encode the positions of individual agents ($N$ matrices), obstacles (one matrix), and objects (one matrix), yielding an $N_c\times R_X\times R_Y$ local observation tensor. The observing agent is always located at the center of the local observation.
\par

Moreover, in sfDA7-X, agent $i$ encodes a $G_X\times G_Y$ {\em positional matrix} that represents its global position, where the entry corresponding to agent $i$'s current node $c_i(t)$ is set to one, and all other entries are zero. While the observation matrices describe the agent's immediate local surroundings, the positional matrix captures its absolute position within the global environment. Consequently, a change in the number of agents $N$ alters the dimensions or structure of the input tensors, necessitating the complete retraining of the agents.
\par

\subsection{Multi-Head Attention}
Self-attention is defined for query, key, and value matrices $Q$, $K$, and $V$, respectively, as~\cite{vaswani2017attention}
\begin{equation}
  \operatorname{Attention}(Q,K,V)=\operatorname{Softmax}\left(\frac{QK^{\mathsf{T}}}{\sqrt{d}}\right)V,
  \label{eq:translation-attention}
\end{equation}
where $d$ denotes the dimension of the queries and keys. Multi-head attention with $h$ attention heads is formulated as
\begin{align}
  \begin{aligned}
    \operatorname{MHA}&(Q,K,V)\\
    &=\operatorname{Concat}(\mathrm{head}_1,\ldots,\mathrm{head}_h) W^O,
  \end{aligned}\\
  \mathrm{head}_l=\operatorname{Attention}(QW_l^Q,KW_l^K,VW_l^V),
\end{align}
where $W_l^Q$, $W_l^K$, $W_l^V$, and $W^O$ are the trainable parameter matrices. The sfDA7-X architecture employs this mechanism to extract features essential for action selection from the given conditional information and local observations.
\par

\begin{figure}[t]
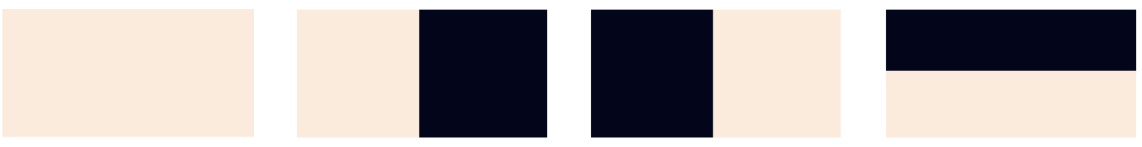

  \dcintroimage{entire.PNG}\hfill
  \dcintroimage{half0.PNG}\hfill
  \dcintroimage{half1.PNG}\hfill
  \dcintroimage{half2.PNG}
  \caption{Examples of DCs (beige: 1; black: 0).}
  \label{fig:translation-dc-example}
\end{figure}

\subsection{Destination Channel}
A DC is a $G_X \times G_Y$ binary matrix representing the specific region in which an agent should work~\cite{motokawa2023strategy}. Figure~\ref{fig:translation-dc-example} shows a representative example; beige cells indicate the region in which the assigned agents are required to collect objects, whereas black cells indicate the region where those agents should not. The matrix elements of the DC corresponding to the beige cells are set to $1$, whereas those corresponding to the black cells are set to $0$. During training, an agent's policy network receives the observation matrix alongside its assigned DC. The agent receives a positive reward $r_o$ when collecting an object within a region where the corresponding DC element is $1$; conversely, it receives no positive reward for collecting an object in a region where the corresponding DC element is $0$. By providing a specific DC shape at execution time, a manager can effectively control the work region at a high level without specifying a detailed action sequence.
\par

\begin{figure}
  \centering
  \includegraphics[width=0.99\columnwidth]{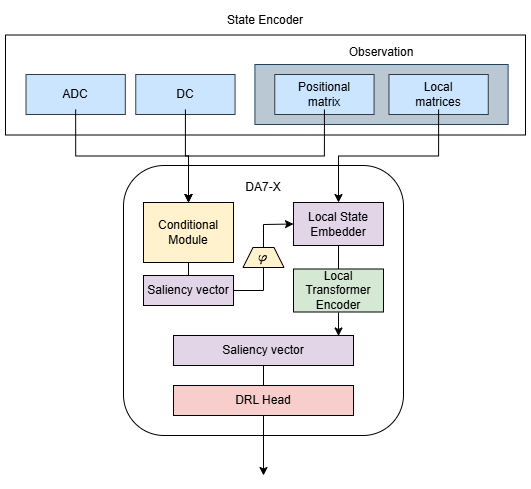}
  \caption{Network Architecture of sfDA7-X.}
  \label{fig:translation-da7}
\end{figure}

\subsection{sfDA7-X and Aggregated DC}
The sfDA7-X architecture~\cite{takahagi2025strategy} extends sfDA6-X~\cite{motokawa2023strategy} to allow agents to consider other agents' DCs alongside their own. The set of all DCs is denoted by $\DCSet$. Figure~\ref{fig:translation-da7} shows sfDA7-X's network architecture, comprising a {\em conditional module} (CM), local Transformer encoder, and DRL head specified by "X" (e.g., sfDA7-IQN, where IQN~\cite{dabney2018implicit} refers to an {\em implicit quantile network}). The CM receives conditional-state matrices: the positional matrix, the own DC, and an {\em aggregated DC} (ADC) encoding other agents' DCs. Each matrix is divided into patches, embedded, and processed by a Transformer encoder in a separate submodule. First-token features from these submodules are integrated into a saliency vector that represents conditional importance. This saliency vector and patched local observation are processed by the local Transformer encoder, after which the DRL head calculates the action values while reusing the saliency vector.
\par

During training, a DC from $\DCSet$ is assigned to each agent in each episode. Let $\DCSet_{e,i}$ be the set of DCs assigned to agents other than agent $i$ in the $e$-th episode. Then, the aggregated DC provided to agent $i$, denoted by $M_i^{\mathrm{ADC}}$, is defined as
\begin{equation}
  M_i^{\mathrm{ADC}} = \sum_{D \in \DCSet_{e,i}} D.
  \label{eq:translation-adc}
\end{equation}
Thus, the ADC is generated by aggregating the multiple DCs assigned to the other agents into a single $G_X \times G_Y$ matrix, making its spatial size independent of the number of agents. If multiple DCs assigned to other agents overlap on the same cell, the corresponding ADC element equals the number of agents assigned to that cell. Note that when no DC is assigned to agent $i$, an all-ones matrix is provided by default, allowing agent $i$ to operate across the entire environment.
\par

In sfDA7-X, each agent inputs its assigned DC and the ADC to the CM as conditional information. Consequently, each agent can infer the work regions of other agents from their instructions while simultaneously considering its own instruction. In particular, an agent not assigned to a specific region learns to avoid overlapping with the work regions of other agents and to complement regions left uninstructed. Generating the ADC at execution time assumes that the system has access to the DCs assigned by the manager to all participating agents.
\par

\subsection{Issue of Training Cost}
Because each agent is assigned a DC in an episode during the training phase, the number of possible combinations of DCs assigned to agents increases exponentially with both the team size and number of DC types. For example, with five candidate DCs, independently assigning one DC to each of $N$ agents produces $5^N$ possible assignments. Table~\ref{tab:translation-training-scale} shows this combinatorial growth, together with the theoretical minimum number of episodes required to experience every assignment at least once with a probability of at least 99\% (referred to as $T_{99}$ in Table~\ref{tab:translation-training-scale}). Therefore, repeated training whenever the number of agents changes is prohibitively costly; this motivates us to reuse a policy trained on a single (e.g., smaller) team size across different team sizes, either without further training or as an initialized state for warm-start retraining.
\par

\begin{table}
  \caption{Combination of DCs and Training Episodes.}
  \label{tab:translation-training-scale}
  \centering
  \tabcolsep=5pt
  \small
  \begin{tabular}{lrr}
    \toprule
    $N$ & $5^N$ & Required Episodes ($T_{99}$) \\
    \midrule
    3 & 125 & 1,174 \\
    4 & 625 & 6,894 \\
    5 & 3,125 & 39,517 \\
    6 & 15,625 & 222,756 \\
    8 & 390,625 & 6,826,419 \\
    \bottomrule
  \end{tabular}
\end{table}

\section{\uppercase{Proposed Method}}\label{sec:translation-method}
\subsection{Overview}
In this study, we extended sfDA7-X by introducing an agent-count-independent observation architecture: {\em strategy-following distributed attentional actor architecture with aggregated DC and agent-aggregated observations} (sfDA8-X). In sfDA8-X, the observation representation is modified to aggregate other agents' observations into a single channel, rendering the input representation independent of the agent count. This design eliminates observation-shape incompatibilities, allowing network structures and trained parameters obtained from a specific team size to be directly deployed across other team sizes via policy replication without additional training. Although all agents share identical parameters, their inputs---comprising local observations, individual DCs, and ADCs---diverge, enabling them to execute distinct actions tailored to their respective contexts.
\par

\subsection{Replication of Trained Policy}
Let $\AgentCountTrain$ denote the number of agents during base training, and let $\theta_i^{\mathrm{train}}$ denote the policy network parameters (i.e., weights) of trained agent $i$. For policy replication, we choose a single source agent $c \in \{0, \ldots, \AgentCountTrain-1\}$ after base training and assign its trained parameters to $\AgentCountEval$ agents, where $\AgentCountEval > 0$ is the target team size and may differ from $\AgentCountTrain$.
\begin{equation}
  \theta_i^{\mathrm{eval}} = \theta_c^{\mathrm{train}},
  \textrm{ for } i = 0, \ldots, \AgentCountEval-1.
\label{eq:translation-replication}
\end{equation}
We refer to the phase in which the target team is trained and evaluated using the replicated policy as the evaluation phase. In zero-shot sfDA8-X, no additional learning or fine-tuning is performed after policy replication in Eq.~\eqref{eq:translation-replication}. In contrast, sfDA8-X-WR, defined in Section~\ref{sec:translation-experiments}, uses the parameters to initialize retraining at the target team size. This design reduces the set of trained policies that must be managed for the evaluation team to a single policy type. However, it does not preserve the heterogeneity among agent-specific policies developed during training; therefore, the choice of source $c$ can affect evaluation performance.
\par

\subsection{Agent-Count-Independent Observation Matrix}
\label{sec:agent-count-independent-observation}
To directly reuse the trained parameters $\theta_c^{\mathrm{train}}$ when the team size changes, the positions of other agents are not represented separately in identity-specific channels but are instead aggregated into a single matrix. For this purpose, agent $i$ does not distinguish other agents based on their individual identities or IDs. Consequently, sfDA8-X uses only three observation channels regardless of whether $\AgentCountTrain = \AgentCountEval$ or $\AgentCountTrain \neq \AgentCountEval$: one channel representing all nearby agents, one representing obstacles, and one representing objects to be collected. Because the observing agent is always located at the center of its local observation grid, a separate channel for the agent is unnecessary. Thus, whereas the conventional sfDA7-X requires $N + 2$ channels (where $N$ is the number of agents), sfDA8-X maintains a constant number of channels, $N_c = 3$.
\par

Although this representation modification ensures compatibility for both the network structure and trained parameter $\theta_c^{\mathrm{train}}$ under varying team sizes, it discards the identity-specific positional information of other agents, which could potentially hinder coordinated behavior during training. However, in exchange for this aggregation, a single policy architecture can be directly deployed across different team sizes. As demonstrated in our evaluations, sfDA8-X achieved training performance comparable to that of sfDA7-X in terms of both total reward and number of collected objects, demonstrating that identity aggregation does not degrade the policy's learning capability. However, input-shape compatibility alone does not guarantee coordinated performance across different team sizes, making experimental evaluation at target team sizes essential.
\par

\section{\uppercase{Experimental Evaluation and Discussion}}
\label{sec:translation-experiments}
\subsection{Experimental setup}
\begin{figure}
  \centering
  \includegraphics[width=\columnwidth]{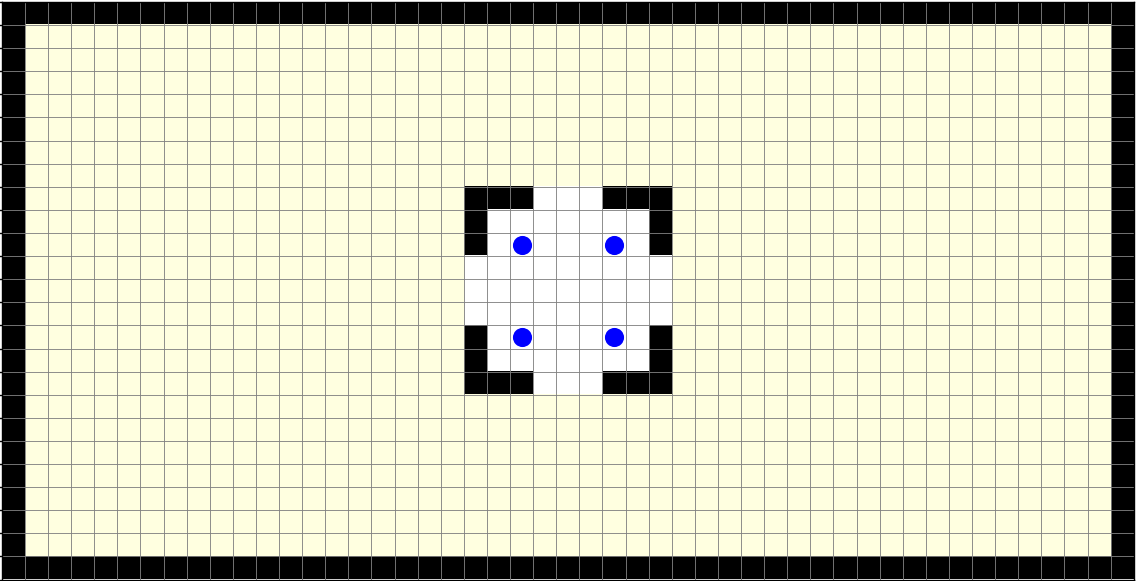}
  \caption{Environment of Object Collection Game.}
  \label{fig:translation-map}
\end{figure}

Figure~\ref{fig:translation-map} shows the $49\times25$ grid environment for the object collection game, where black cells represent walls, beige cells represent object spawn regions, and blue circles represent example initial positions for a four-agent team. The agents were randomly assigned to these positions without duplication in each episode. Table~\ref{tab:translation-positions} lists the initial positions for different team sizes. Each episode ($H=200$ steps) began with 80 objects scattered in the beige regions. When an agent collected an object, it vanished, and another spawned elsewhere in the beige region. The local observation size was specified as $R_X=R_Y=7$. The rewards for object collection, collision (with walls or agents), and other movements were $r_o=1$, $r_c=-1$, and $r_m=0$, respectively.
\par

The DRL head used an IQN~\cite{dabney2018implicit}, yielding the instances sfDA7-IQN and sfDA8-IQN. While sfDA7-IQN used $N_c=N+2$ observation channels by representing each agent separately, sfDA8-IQN maintained $N_c=3$. Hyperparameters included a learning rate of $10^{-3}$, discount factor $\gamma=0.9$, Adam $\varepsilon=10^{-8}$, and batch size 32. The $\varepsilon$-greedy exploration rate decayed from $1.0$ to $0.05$ with a factor of $0.998$
\par

\begin{figure}
  \centering
  \dcimage{entire.PNG}\hfill
  \dcimage{half0.PNG}\hfill
  \dcimage{half1.PNG}\hfill
  \dcimage{half2.PNG}\hfill
  \dcimage{half3.PNG}
  \caption{DC Types used during Training.}
  \label{fig:translation-training-dcs}
\end{figure}

\begin{table}
  \caption{Initial Positions for Each Team Size.}
  \label{tab:translation-positions}
  \centering
  \small
  \begin{tabular}{c>{\centering\arraybackslash}p{5.3cm}}
    \toprule
    Agents & Candidate positions \\
    \midrule
    3 & $(-2,2),(2,2),(0,-2)$ \\
    4 & $(\pm2,\pm2)$ \\
    5 & $(\pm2,\pm2),(0,0)$ \\
    6 & $(\pm2,2),(\pm2,0),(\pm2,-2)$ \\
    8 & $\{(x,y)\mid x,y\in\{-2,0,2\},(x,y)\neq(0,0)\}$ \\
    \bottomrule
  \end{tabular}
\end{table}

\begin{figure*}
\centering
\begin{minipage}{0.45\textwidth}
  \includegraphics[width=\textwidth]{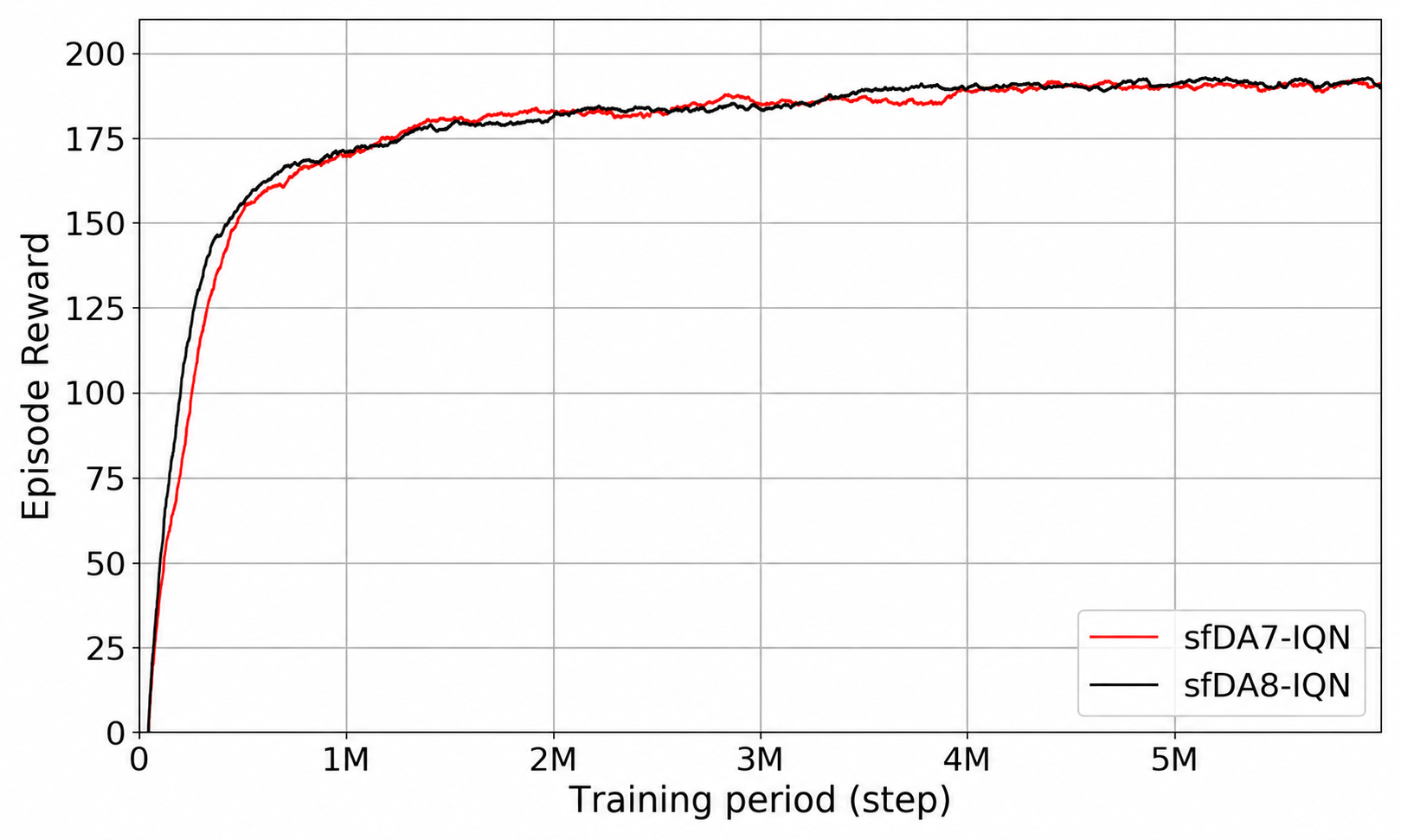}
  \subcaption{Episode reward}
\end{minipage}
\hfill
\begin{minipage}{0.45\textwidth}
  \includegraphics[width=\textwidth]{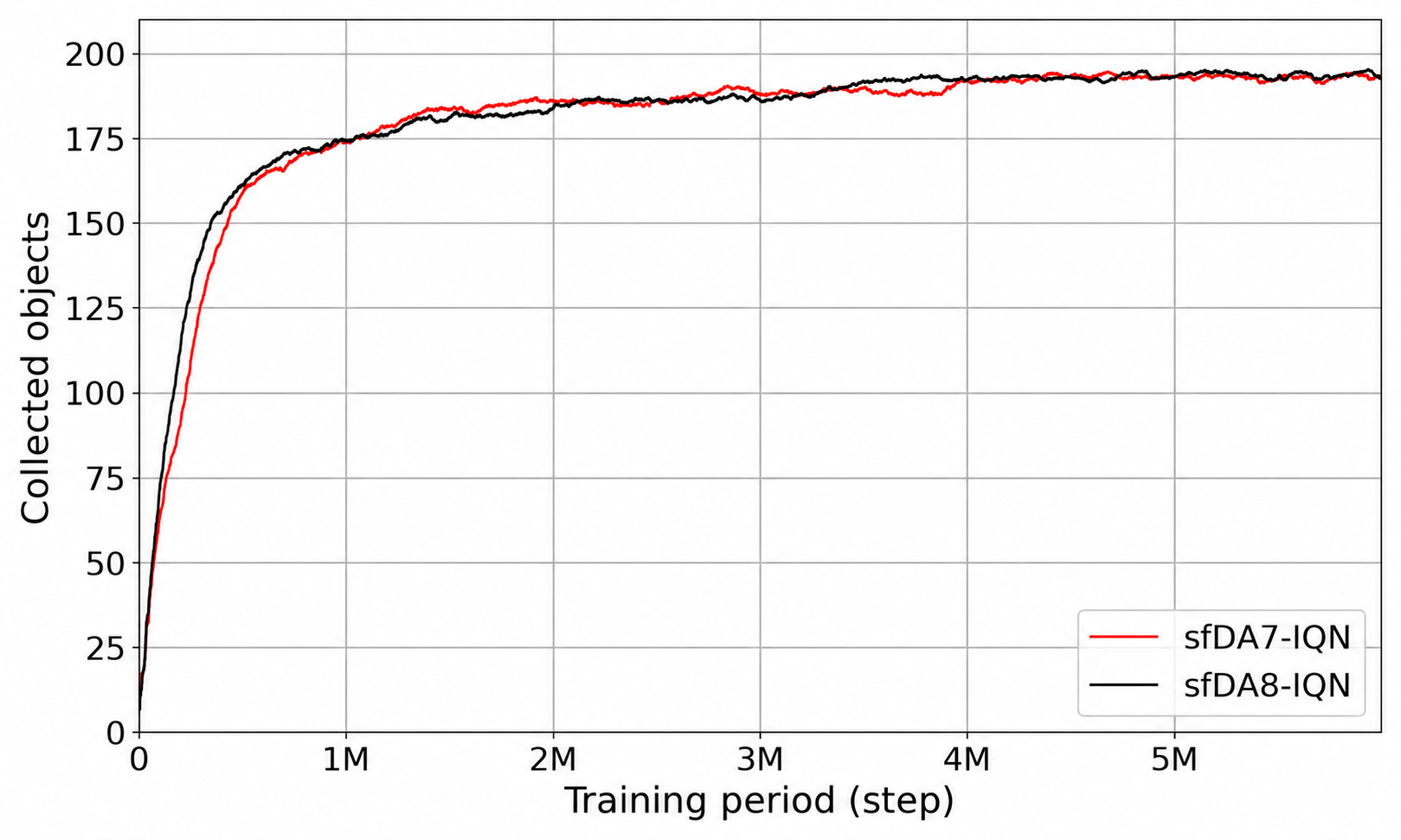}
  \subcaption{Number of collected objects}
\end{minipage}
\caption{Learning Curves of sfDA7-IQN and sfDA8-IQN with Four Agents.}
\label{fig:translation-da7-da8-learning}
\end{figure*}

\subsection{Comparison of Learning Performance}
We compared the learning curves of sfDA7-IQN and sfDA8-IQN with four agents in Experiment 1 (Exp.~1) to examine the effect of replacing identity-specific observations with the proposed three-channel aggregated observation. Note that the agents learn their policies independently. The five DCs used during training for both frameworks are shown in Figure~\ref{fig:translation-training-dcs}: the entire DC and the left, right, lower, and upper half-DCs. An agent assigned the entire DC can collect objects anywhere in the beige regions, whereas an agent assigned a half-DC is restricted to collecting objects within that specific half of the beige regions. In each episode, one of the five DCs was selected with equal probability and independently assigned to each agent.
\par

Figure~\ref{fig:translation-da7-da8-learning} shows the results of Exp.~1. The total reward and the number of collected objects converged to almost identical levels for both models, suggesting that using the agent-count-independent observation matrix did not impair learning performance.
\par

\begin{figure}
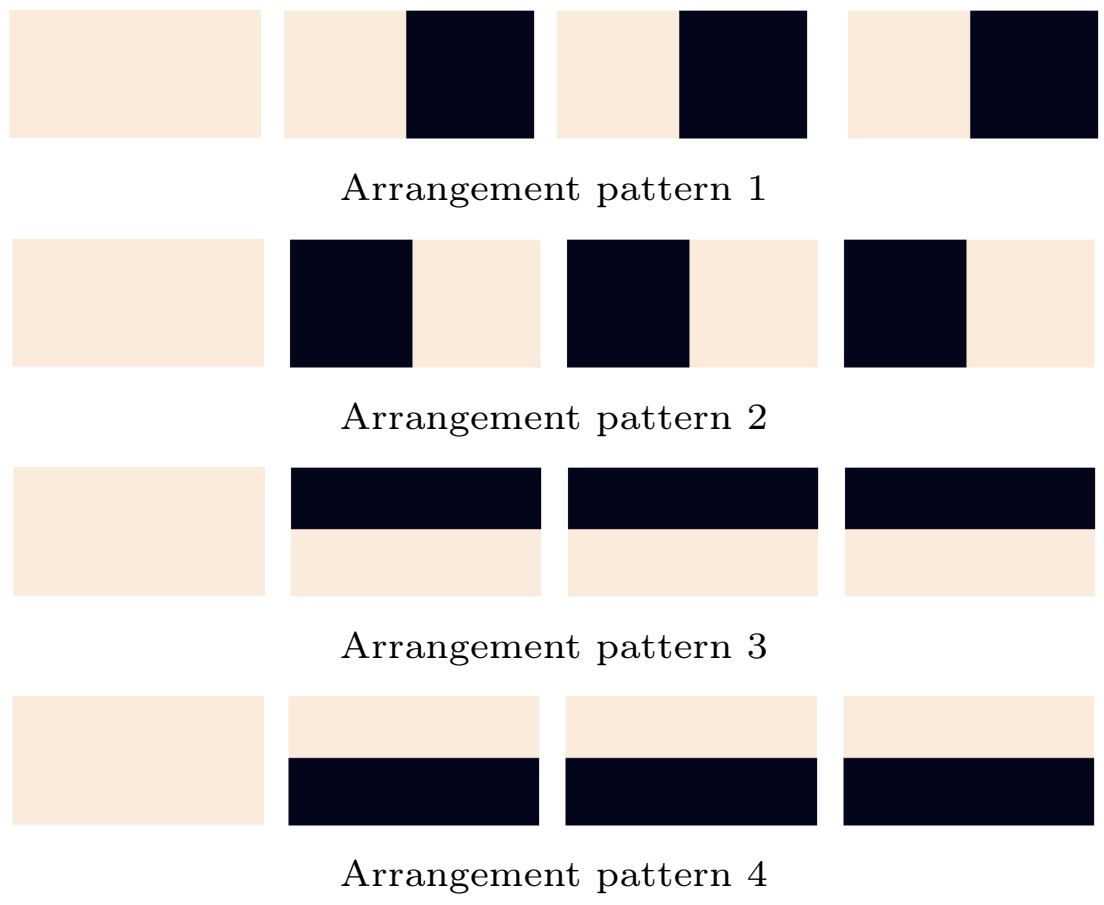

\centering
\dcevalimage{entire.PNG}\hfil\dcevalimage{half0.PNG}\hfil\dcevalimage{half0.PNG}
\hfil\dcevalimage{half0.PNG}\\
{\scriptsize Arrangement pattern 1}\\[5pt]
\dcevalimage{entire.PNG}\hfil\dcevalimage{half1.PNG}\hfil\dcevalimage{half1.PNG}\hfil\dcevalimage{half1.PNG}\\
{\scriptsize Arrangement pattern 2}\\[5pt]
\dcevalimage{entire.PNG}\hfil\dcevalimage{half2.PNG}\hfil\dcevalimage{half2.PNG}\hfil\dcevalimage{half2.PNG}\\
{\scriptsize Arrangement pattern 3}\\[5pt]
\dcevalimage{entire.PNG}\hfil\dcevalimage{half3.PNG}\hfil\dcevalimage{half3.PNG}\hfil\dcevalimage{half3.PNG}\\
{\scriptsize Arrangement pattern 4}\\[5pt]
\caption{Examples of DC Arrangement Patterns.}
\label{fig:translation-eval-dcs}
\end{figure}

\subsection{Zero-Shot Cross-Team-Size Evaluation}\label{sec:Zero-Shot} 
The objective of the second experiment (Exp.~2) was to transfer a policy selected from the four policies learned by the four-agent team to teams of three, four, and five agents in a zero-shot manner, comparing their performance with that achieved when each team learned its policies from scratch. Note that we only compare the performance of sfDA8-IQN-IP and sfDA8-IQN-RP, as sfDA7-IQN cannot be applied to the zero-shot cross-team-size situations.
\par

We compared two types of sfDA8-IQN conditions in Exp.~2. In sfDA8-IQN using individual policies (sfDA8-IQN-IP), agents in teams of three to five agents used their own policies trained with randomly assigned DCs for $10,000$ (three agents), $30,000$ (four agents), and $50,000$ (five agents) episodes. These values were determined based on the data listed in Table~\ref{tab:translation-training-scale}. Note that $30,000$ episodes correspond to $6\text{M}$ steps, which is identical to the training length in Exp.~1. In contrast, in sfDA8-IQN using a replicated policy (sfDA8-IQN-RP), one of the four policies trained under sfDA8-IQN-IP for $30,000$ episodes was selected, and its network parameters were copied to the networks of all agents. Thus, all agents shared identical network parameters, but their local observations, individual DCs, and ADCs could differ; consequently, their actions differed according to their inputs. The selected policy in a given evaluation is hereafter referred to as the \emph{source policy}. 
\par

When evaluating the performance of sfDA8-IQN-IP and sfDA8-IQN-RP, we used the four DC arrangement patterns shown in Fig.~\ref{fig:translation-eval-dcs} as an example for a four-agent team: one entire DC was assigned to one agent and one half-DC was assigned to all the remaining agents. For instance, in Pattern~1, the team treated the left-half region as the \emph{key region} for collecting objects. We then evaluated the total rewards and number of collected objects over $1,000$ episodes for all DC arrangement patterns; for instance, this yielded $12$ ($=3 \times 4$) evaluations for the three-agent team. The reason we evaluated sfDA8-IQN using these DC patterns was to confirm that sfDA8-IQN retained the key characteristic of sfDA7-IQN, namely, that an agent given an entire DC can cover areas not specifically instructed to be covered by other agents.
\par

\begin{table}
\caption{Episode Rewards and Collected Objects.}
\label{tab:translation-average-results}
\centering
\small
\tabcolsep=2pt
\begin{tabular}{clrrr}
\toprule
$N$ & Metric & sfDA8-IQN-IP & sfDA8-IQN-RP & Diff.(\%) \\
\midrule
3 & Reward & 103.118 & 116.068 & $+12.56$ \\
 & Objects & 117.845 & 117.618 & $-0.19$ \\[3pt]
4 & Reward & 141.536 & 131.533 & $-7.07$ \\
 & Objects & 143.814 & 134.130 & $-6.73$ \\[3pt]
5 & Reward & 145.556 & 140.311 & $-3.60$ \\
 & Objects & 149.951 & 145.785 & $-2.78$ \\
\bottomrule
\end{tabular}
\end{table}

\begin{figure*}
\centering
\begin{minipage}{0.48\textwidth}
  \centering
  \includegraphics[width=0.90\textwidth]{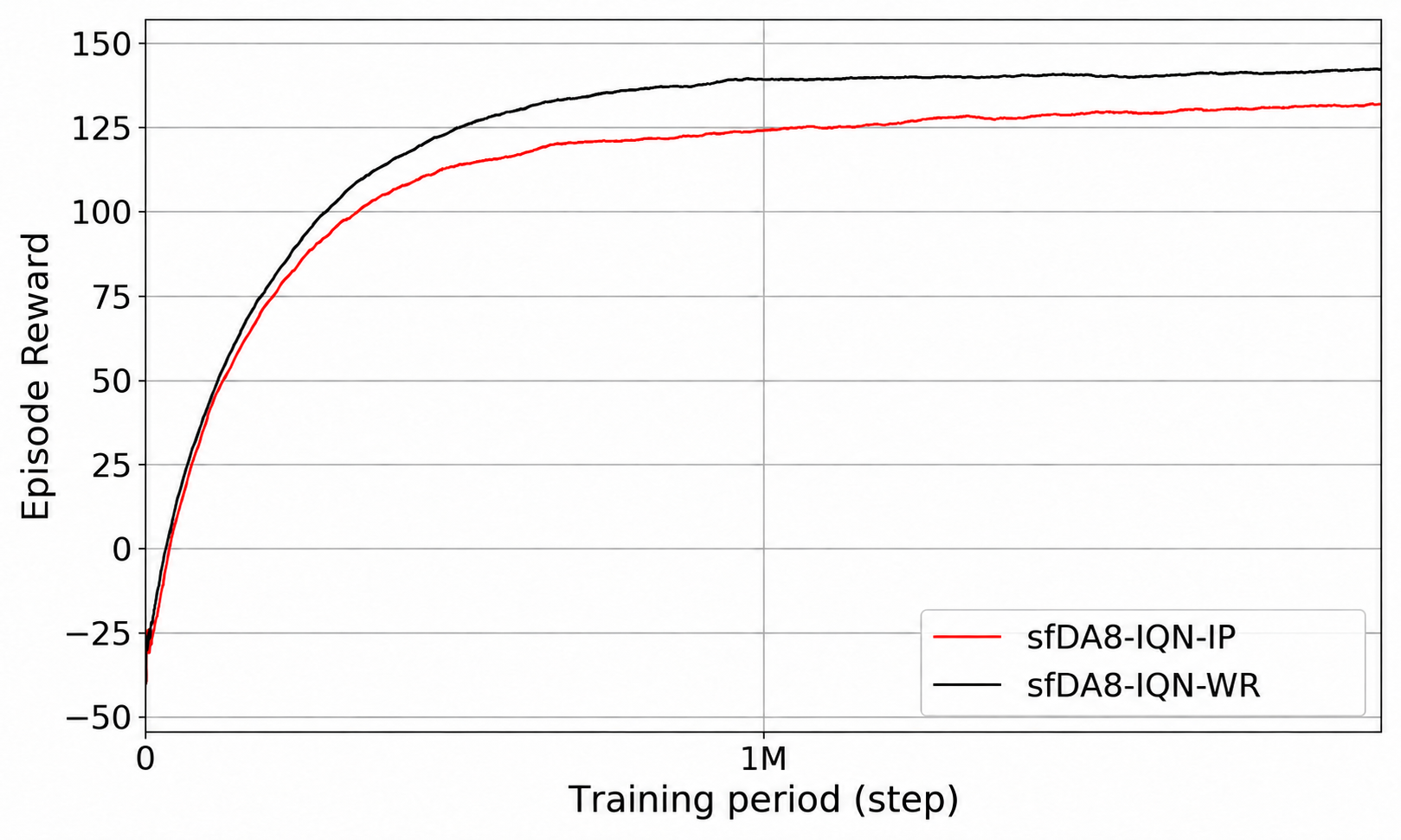}
\subcaption{Episode reward (three-agent team)}
\end{minipage}
\hfill
\begin{minipage}{0.48\textwidth}
  \centering
  \includegraphics[width=0.90\textwidth]{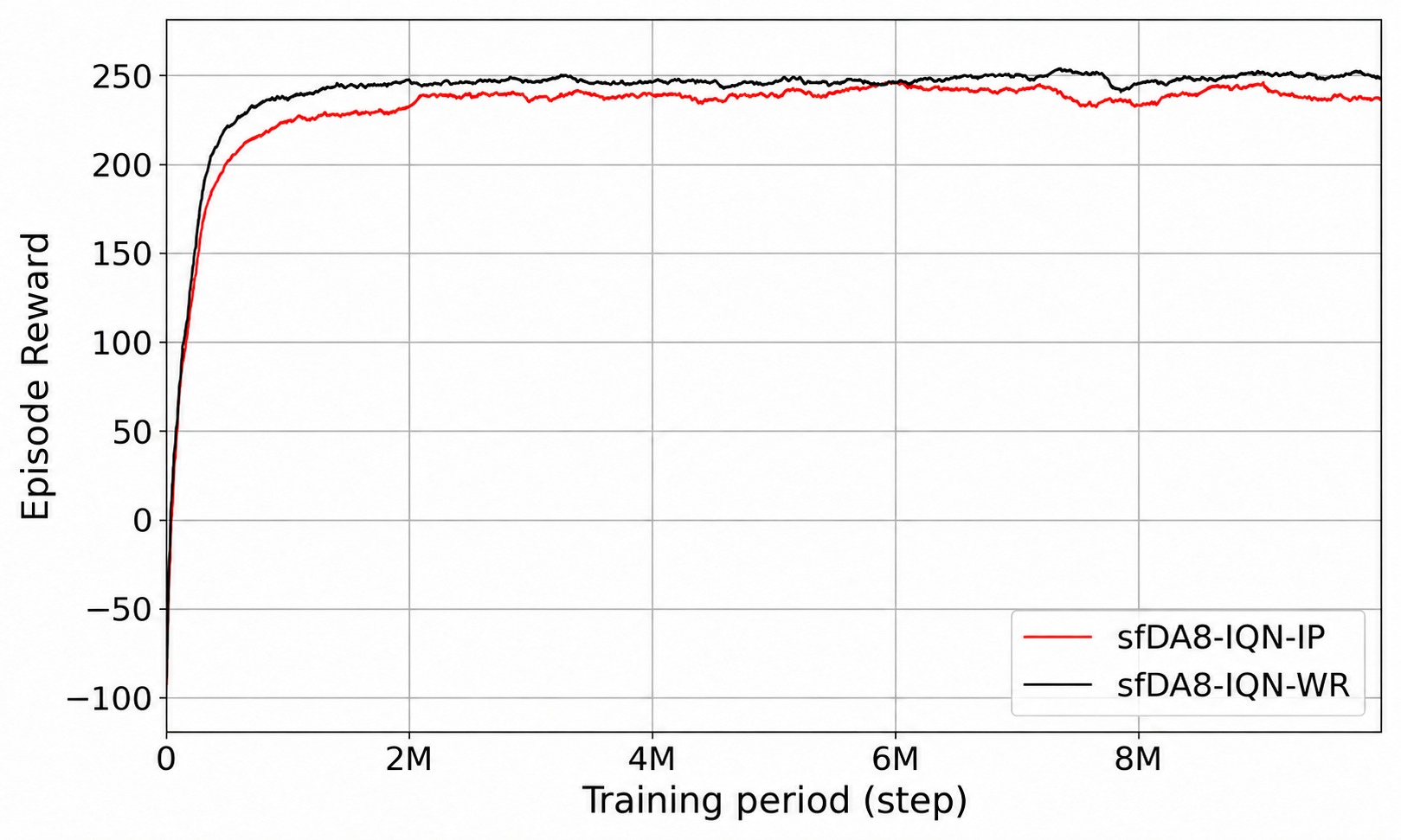}
\subcaption{Episode reward (five-agent team)}
\end{minipage}
\\[10pt]
\begin{minipage}{0.48\textwidth}
  \centering
  \includegraphics[width=0.90\textwidth]{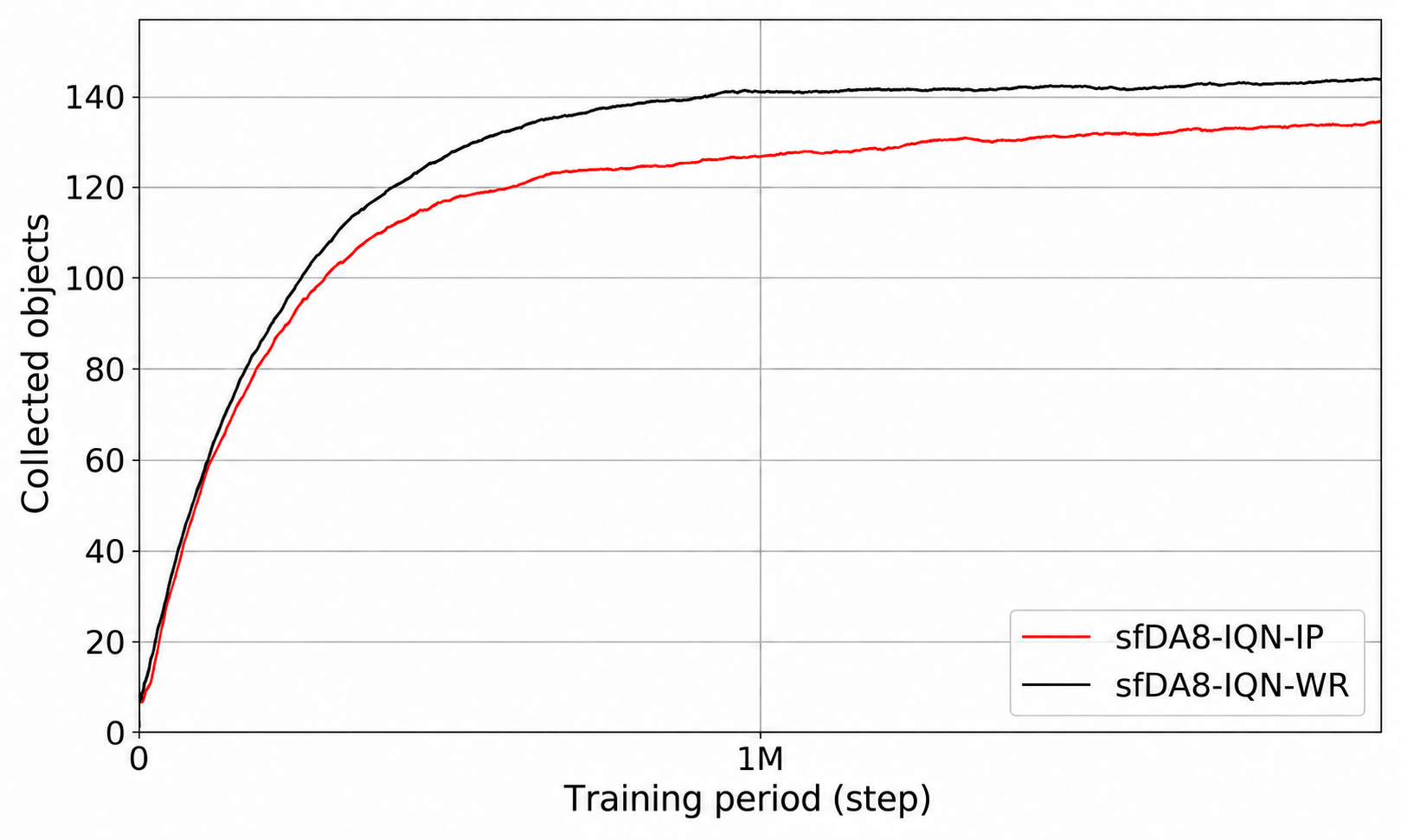}
  \subcaption{Number of collected objects (three-agent team)} 
\end{minipage}
\hfill
\begin{minipage}{0.48\textwidth}
  \centering
  \includegraphics[width=0.90\textwidth]{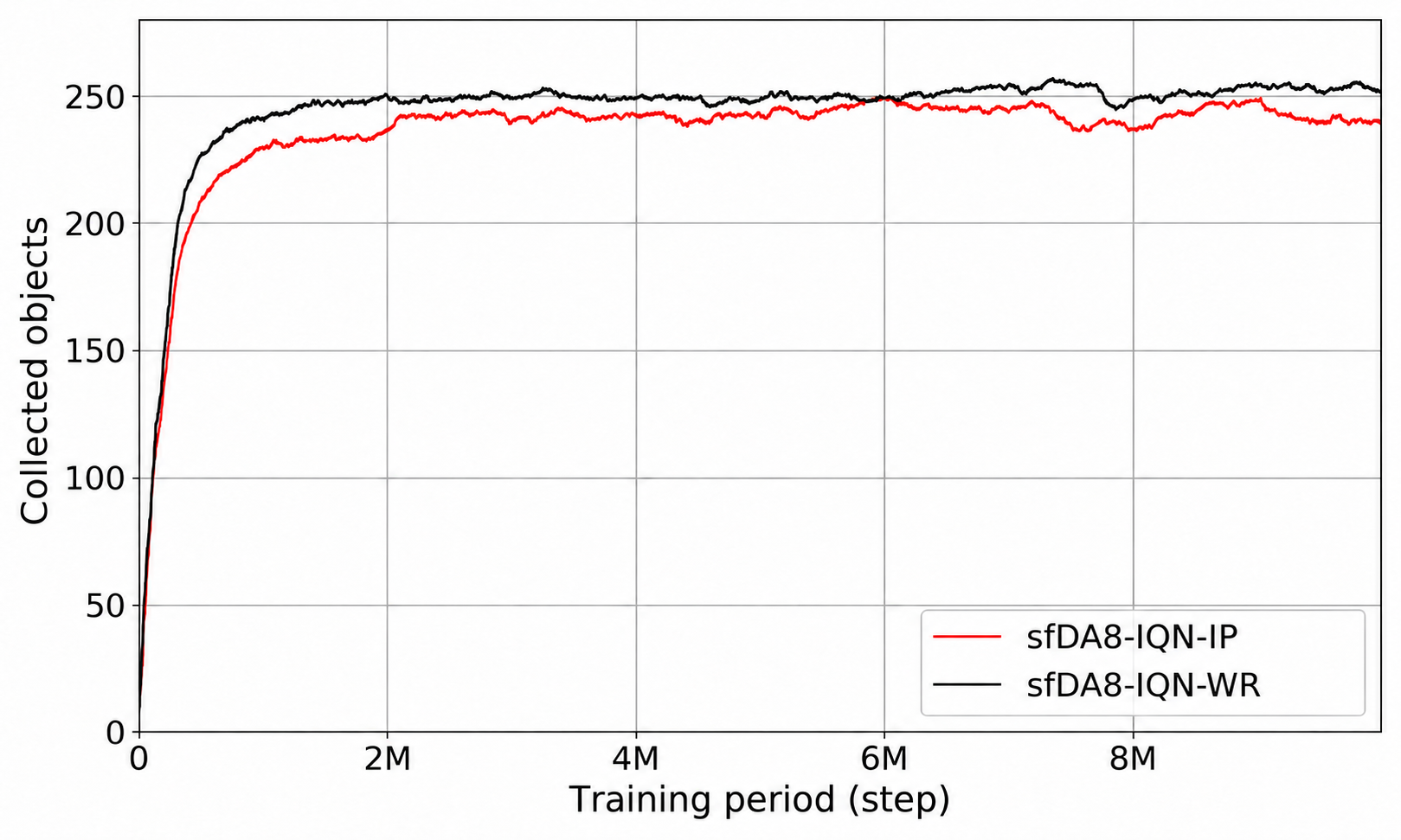}
  \subcaption{Number of collected objects (five-agent team)} 
\end{minipage}
\\[10pt]
\caption{Training Curves of sfDA8-IQN-WR and sfDA8-IQN-IP.}
\label{fig:translation-wr-learning-curves}
\end{figure*}

The experimental results of Exp.~2 are summarized in Table~\ref{tab:translation-average-results}, which lists the average episode reward, average number of collected objects, and percentage differences between them. In the case of the four-agent team---a reference experiment that isolated policy replication without changing team size---the reward and number of collected objects for sfDA8-IQN-RP decreased by $7.07\%$ and $6.73\%$, respectively, compared with those for sfDA8-IQN-IP. This difference can be attributed to reducing diverse individual policies to a single policy, which aligns with intuition.
\par

Counterintuitively, in the three-agent team, sfDA8-IQN-RP increased the reward by $12.56\%$ relative to that of sfDA8-IQN-IP, while the mean number of collected objects showed a negligible change of $-0.19\%$. Although object collection performance under the DC instructions remained almost identical, the reward values increased significantly. This suggests that while $10,000$ episodes (the training duration for sfDA8-IQN-IP) were sufficient for collecting objects in specified regions, the RP agents---trained for $30,000$ episodes---achieved superior coordinated behavior, such as more effective collision avoidance. In the five-agent team, the reward and number of collected objects for sfDA8-IQN-RP decreased by only $3.60\%$ and $2.78\%$, respectively. Although the sfDA8-IQN-IP agents were trained longer ($50,000$ episodes) than the RP agents ($30,000$ episodes), this performance gap remains remarkably small. Considering the substantially higher computational cost of training from scratch, these results demonstrate that zero-shot policy replication is highly effective and provides a solid foundation for warm-start retraining across different team sizes.
\par

\begin{table*}
\caption{Warm-Start sfDA8-IQN-WR Performance at Checkpoints.}
\label{tab:warm-start-results}
\centering
\normalsize
\begin{tabular}{crrrrr}
\toprule
Team size ($N$) & Checkpoint & Reward & Difference (\%) & Objects & Difference (\%) \\ 
\midrule
3 & 3,000 & 125.911 & $+22.10$ & 127.059 & $+7.82$ \\
3 & 10,000 & 133.307 & $+29.28$ & 133.769 & $+13.51$ \\
5 & 5,000 & 149.139 & $+2.46$ & 153.445 & $+2.33$ \\
5 & 10,000 & 157.631 & $+8.30$ & 159.257 & $+6.21$ \\
5 & 20,000 & 161.357 & $+10.86$ & 163.213 & $+8.84$ \\
5 & 50,000 & 149.990 & $+3.05$ & 159.521 & $+6.38$ \\
\bottomrule
\end{tabular}
\end{table*}

\subsection{Retraining with Warm-Start Replicated Policies}
In the third experiment (Exp.~3), we conducted warm-start retraining following policy replication. The source policy---trained in four-agent teams and exhibiting the highest average performance in Exp.~2---was selected as the base policy. Its network parameters $\theta_c^{\mathrm{train}}$ were then copied across the initial replicated policy networks of all agents in sfDA8-IQN; hereafter, this warm-start retraining framework with a replicated policy is denoted as sfDA8-IQN-WR. We compared the training performance of sfDA8-IQN-WR with that of sfDA8-IQN-IP. We set evaluation checkpoints at $3,000$ and $10,000$ episodes for the three-agent team and at $5,000$, $10,000$, $20,000$, and $50,000$ episodes for the five-agent team. At each checkpoint, training was interleaved with an evaluation phase comprising $1,000$ test episodes without exploration ($\varepsilon=0$) to record rewards and collected objects for both sfDA8-IQN-IP and sfDA8-IQN-WR.
\par

The resulting learning curves are plotted in Fig.~\ref{fig:translation-wr-learning-curves}. This figure clearly indicates that the warm-start replicated policy (sfDA8-IQN-WR) consistently outperformed sfDA8-IQN-IP. Note that sfDA8-IQN-WR initially recorded negative rewards despite inheriting network parameters from the trained source policy; this was due to resetting the exploration rate to $\varepsilon=1.0$ at the start of retraining.
\par

Table~\ref{tab:warm-start-results} lists the performance metrics at each checkpoint, where ``Difference'' represents the improvement ratio of sfDA8-IQN-WR relative to the final performance of sfDA8-IQN-IP (i.e., at $10,000$ episodes for three-agent teams and at $50,000$ episodes for five-agent teams), with positive values indicating higher rewards or more collected objects. The table demonstrates that for three-agent teams, sfDA8-IQN-WR at $3,000$ episodes already exceeded the final performance of sfDA8-IQN-IP by $+22.10\%$ in reward and $+7.82\%$ in collected objects, with performance further increasing at $10,000$ episodes. Notably, the improvement in total reward was more pronounced than that in object count, indicating that sfDA8-IQN-WR agents acquired superior coordination capabilities---not only collecting objects efficiently but also navigating cautiously to avoid agent--agent collisions.
\par

Similarly, for the five-agent team, sfDA8-IQN-WR at the $5,000$-episode checkpoint already surpassed sfDA8-IQN-IP trained for $50,000$ episodes by $+2.46\%$ in reward and $+2.33\%$ in collected objects. Performance peaked at $20,000$ episodes ($161.357$ reward and $163.213$ objects). In contrast, the earned reward was lower at $50,000$ episodes ($149.990$ reward and $159.521$ objects) than at $20,000$ episodes. This performance degradation indicates potential overfitting resulting from excessive parameter updates on the replicated network.
\par

These results demonstrate that warm-start retraining with replicated policies enabled highly efficient, low-cost policy adaptation when team size changed, achieving superior performance with up to a $90\%$ reduction in required training episodes.
\par

\begin{figure*}
  \centering
  \begin{minipage}[t]{0.41\textwidth}
    \begin{minipage}[t]{0.98\textwidth}
      \centering
      \small{\hfil Agent~0\hfil\hfil\hfil\hfil
        Agent~1\hfil\hfil\hfil\hfil
        Agent~2\hfil}\\
      \includegraphics[width=0.295\linewidth]{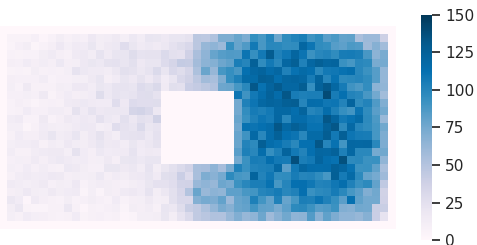}\hfill
      \includegraphics[width=0.295\linewidth]{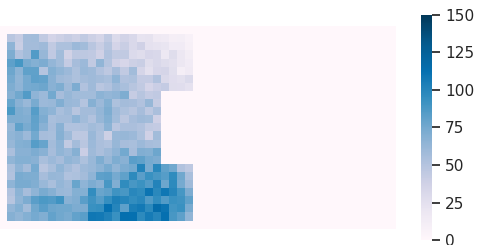}\hfill
      \includegraphics[width=0.295\linewidth]{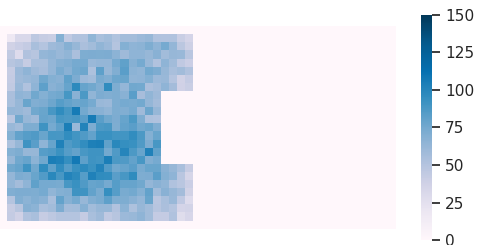}
      \subcaption{sfDA8-IQN-IP}.
    \end{minipage}
    \\
    \begin{minipage}[t]{0.98\textwidth}
      \includegraphics[width=0.295\linewidth]{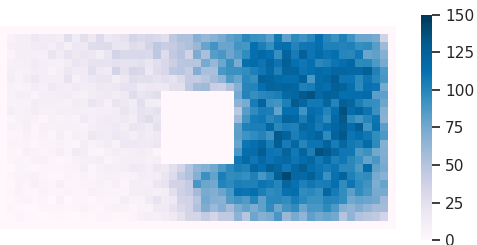}\hfill
      \includegraphics[width=0.295\linewidth]{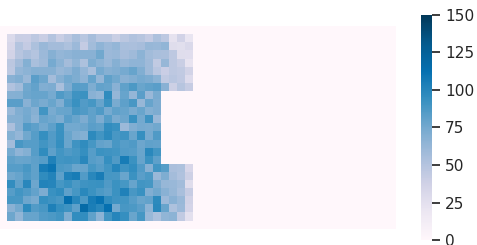}\hfill
      \includegraphics[width=0.295\linewidth]{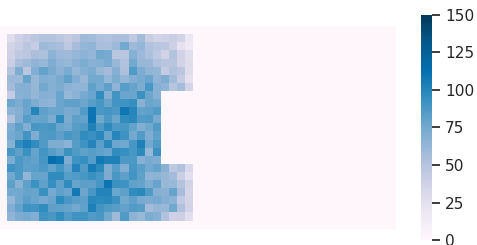}
      \subcaption{sfDA8-IQN-RP}
    \end{minipage}\\[5pt]
    \caption{Collection Locations (3-agent team).}
    \label{fig:translation-heatmap3}
  \end{minipage}
  \hfill
  \begin{minipage}[t]{0.545\textwidth}
    \begin{minipage}[t]{0.98\textwidth}
      \small{\hfil Agent~0\hfil\hfil\hfil
        Agent~1\hfil\hfil\hfil
        Agent~2\hfil\hfil\hfil
        Agent~3\hfil
      }\\
      \includegraphics[width=0.225\linewidth]{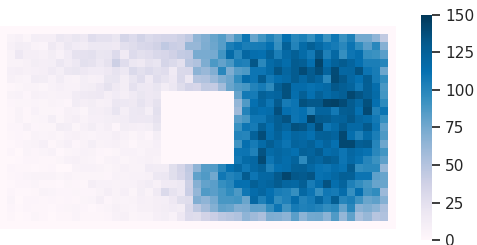}\hfill
      \includegraphics[width=0.225\linewidth]{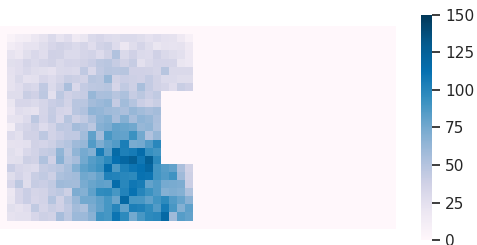}\hfill
      \includegraphics[width=0.225\linewidth]{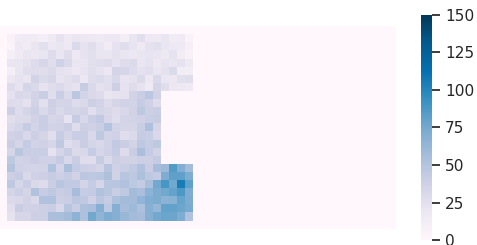}\hfill
      \includegraphics[width=0.225\linewidth]{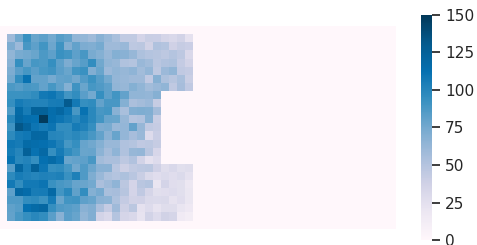}
      \subcaption{sfDA8-IQN-IP}
    \end{minipage}
    \\[5pt]
    \begin{minipage}[t]{0.98\textwidth}
      \includegraphics[width=0.225\linewidth]{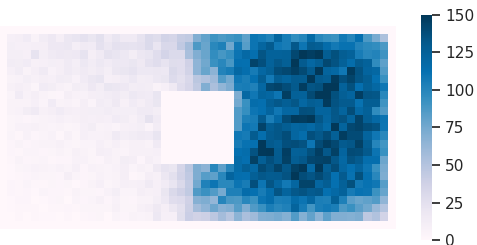}\hfill
      \includegraphics[width=0.225\linewidth]{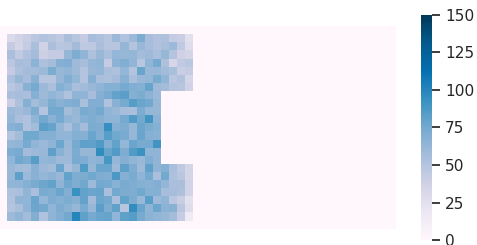}\hfill
      \includegraphics[width=0.225\linewidth]{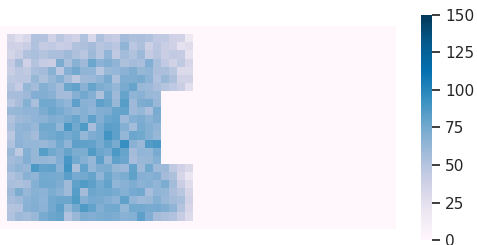}\hfill
      \includegraphics[width=0.225\linewidth]{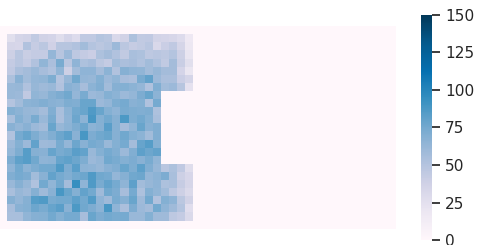}
      \subcaption{sfDA8-IQN-RP}
    \end{minipage}\\[5pt]
    \caption{Collection Locations (4-agent team).}
    \label{fig:translation-heatmap4}
  \end{minipage}
\vspace{10pt}\par
  \centering
  \begin{minipage}{0.80\textwidth}
    \small{\hfil
      Agent~0\hfil\hfil\hfil
      Agent~1\hfil\hfil\hfil
      Agent~2\hfil\hfil\hfil
      Agent~3\hfil\hfil\hfil
      Agent~4\hfil\hfil
    }\\
    \heatmap{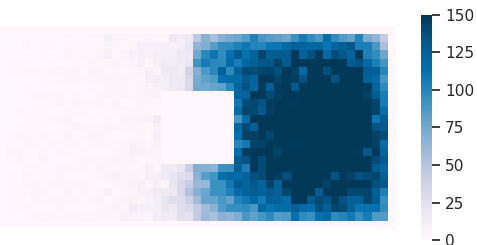}\hfil
    \heatmap{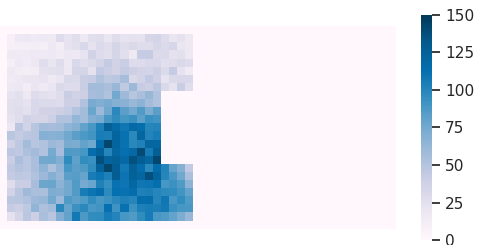}\hfil
    \heatmap{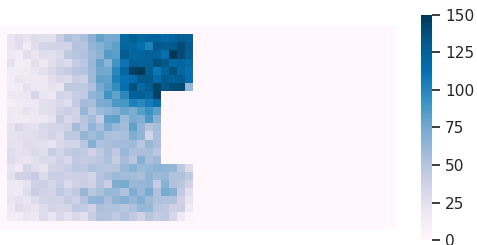}\hfil
    \heatmap{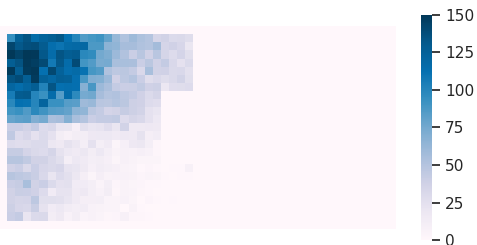}\hfil
    \heatmap{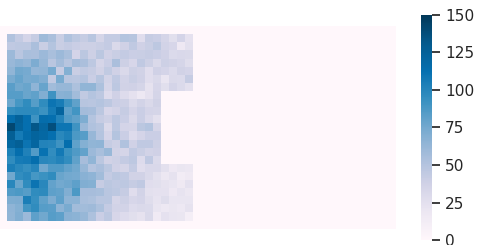}
    \subcaption{sfDA8-IQN-IP}
  \end{minipage}\\[5pt]
  \begin{minipage}{0.80\textwidth}
    \heatmap{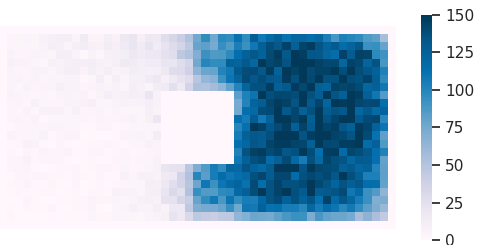}\hfil
    \heatmap{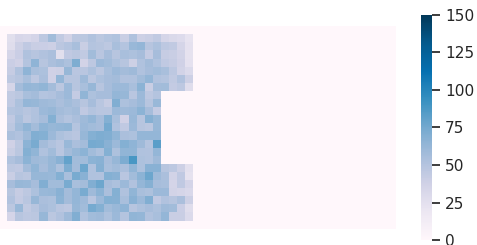}\hfil
    \heatmap{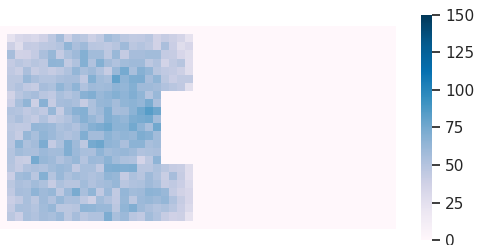}\hfil
    \heatmap{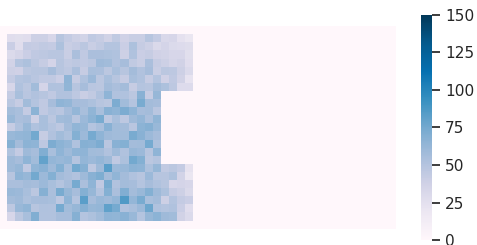}\hfil
    \heatmap{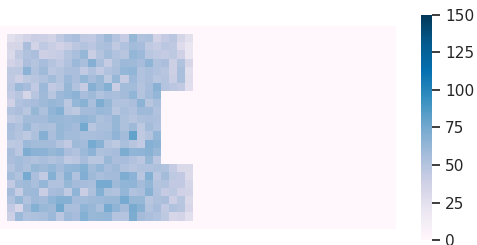}
    \subcaption{sfDA8-IQN-RP}
  \end{minipage}\\[5pt]
  \caption{Collection Locations (5-agent team).}
  \label{fig:translation-heatmap5}
\end{figure*}
\begin{figure*}
  \small{\hspace{4mm}
    Agent~0\hfill
    Agent~1\hfill
    Agent~2\hfill
    Agent~3\hfill
    Agent~4\hfill
    Agent~5\hspace{7mm}
  }\\
  \heatmap{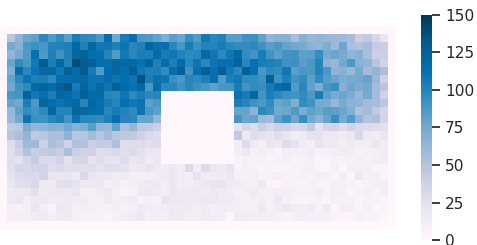}\hfill
  \heatmap{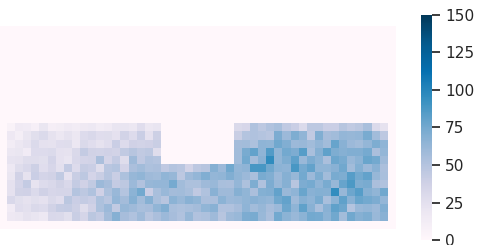}\hfill
  \heatmap{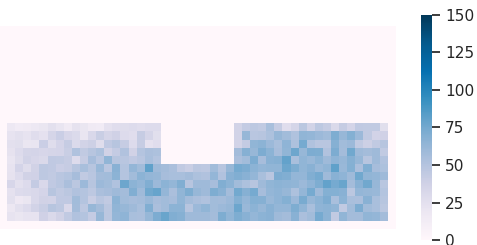}\hfill
  \heatmap{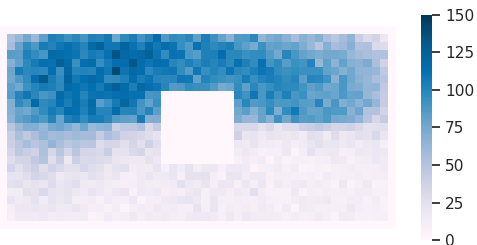}\hfill
  \heatmap{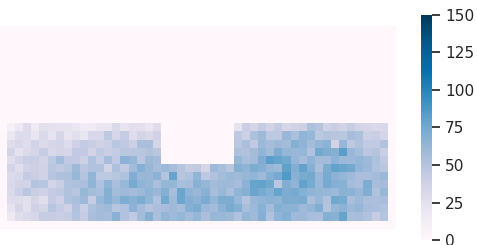}\hfill
  \heatmap{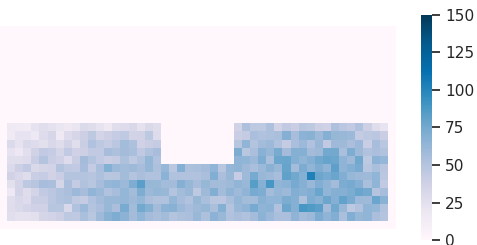}
  \caption{Collection Locations using sfDA8-IQN-RP (6-agent team).}
  \label{fig:translation-heatmap6}
\vspace{10pt}\par
  \centering
  \begin{minipage}{0.70\textwidth}
  \small{\hspace{1mm}
    Agent~0\hfill
    Agent~1\hfill
    Agent~2\hfill
    Agent~3\hspace{3mm}
  }\\
  \heatmap{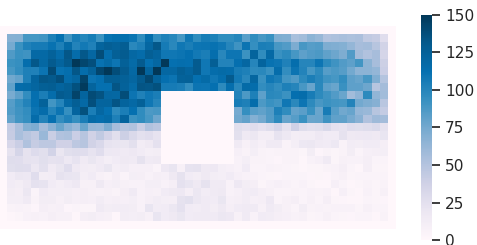}\hfill
  \heatmap{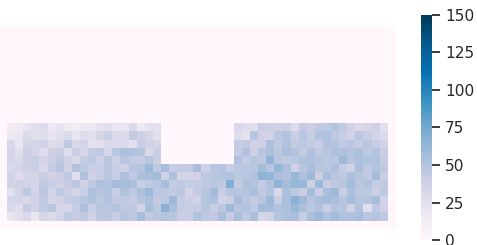}\hfill
  \heatmap{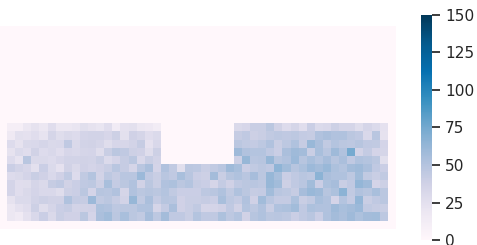}\hfill
  \heatmap{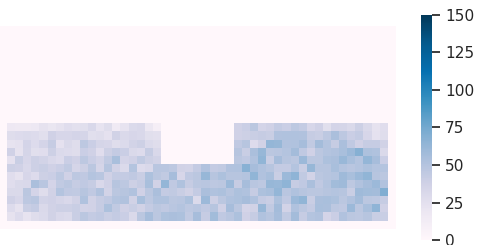}\\[5pt]
    \small{\hspace{1mm}
    Agent~4\hfill
    Agent~5\hfill
    Agent~6\hfill
    Agent~7\hspace{3mm}
  }\\
    \heatmap{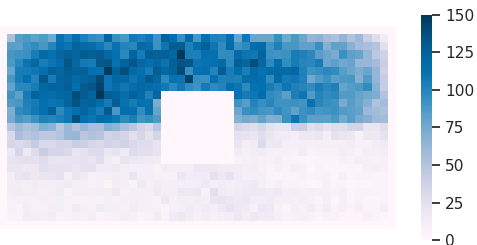}\hfill
    \heatmap{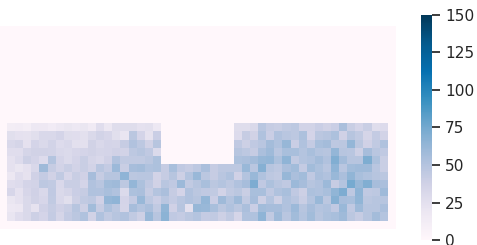}\hfill
    \heatmap{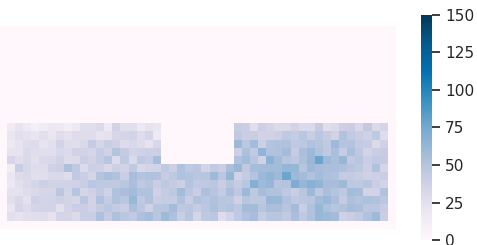}\hfill
    \heatmap{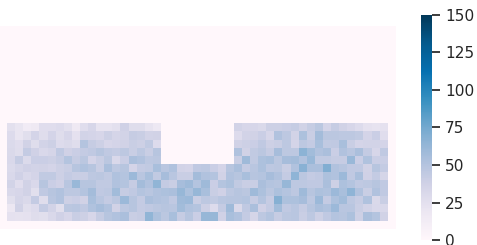}
  \end{minipage}
  \caption{Collection Locations using sfDA8-IQN-RP (8-agent team).}
  \label{fig:translation-heatmap8}
\end{figure*}

\subsection{Analysis of DC Adherence}
We analyzed DC adherence to verify whether agents collected objects in accordance with their assigned DC instructions. The experimental settings were identical to those in Exp.~2.
\par

Figures~\ref{fig:translation-heatmap3}--\ref{fig:translation-heatmap5} show the object collection heatmaps for three-, four-, and five-agent teams. The upper rows in these figures represent sfDA8-IQN-IP, while the lower rows represent sfDA8-IQN-RP for a representative experimental run, where one agent (Agent~0) was assigned the entire DC and the remaining agents received DCs instructing them to operate in the left region of the environment. Color intensity indicates the number of collected objects (darker colors correspond to higher collection counts), with individual color bars displayed on the right-hand side.
\par

In both models, Agent~0 specifically concentrates on gathering objects in the right-hand area, where coverage would otherwise be thin. These figures demonstrate that in sfDA8-IQN-RP, Agent~0 accounted for the DCs assigned to other agents via the ADC, thereby dynamically complementing object collection in the undermanned right region. This adaptive behavior is consistent with that of sfDA8-IQN-IP. However, observing the collection patterns of agents other than Agent~0 reveals a notable difference: in sfDA8-IQN-IP, individual agents tended to specialize in distinct sub-regions within the left area (upper rows), whereas in sfDA8-IQN-RP, agents collected objects uniformly across the left region (lower rows) in strict accordance with their assigned DCs.
\par

We conducted the same analysis for six- and eight-agent teams. In the six-agent team, Agents~0 and 3 were assigned entire DCs, whereas the remaining agents were assigned lower-region DCs. In the eight-agent team, Agents~0 and 4 were assigned entire DCs, whereas the remaining agents were assigned lower-region DCs. The results for sfDA8-IQN-RP are shown in Figs.~\ref{fig:translation-heatmap6} and \ref{fig:translation-heatmap8}. These figures further confirm that the agents assigned entire DCs leveraged their ADCs to focus on supplementing the undermanned upper-half region, while the other agents acted in accordance with their assigned DC instructions in the lower-half region. This phenomenon reflects the interplay between policy diversity and homogeneity, providing empirical support for the discussion in Section~\ref{sec:Zero-Shot}.
\par

\subsection{Discussion}
The experimental results of this study demonstrate that sfDA8-X enables high-quality cooperative behavior across teams of varying sizes by transferring pre-trained policy parameters to all agents. This approach either eliminates the need to train new teams from scratch or significantly reduces computational overhead when used as a warm-start initialization for retraining. Even when agents share an identical replicated policy, they do not exhibit identical behaviors; because each agent receives distinct inputs---comprising its local observation, its uniquely assigned DC, and the collective ADC---the shared policy generates asymmetric, complementary actions that effectively maintain team coordination.
\par

Meanwhile, a practical consideration arises regarding policy selection: in our experiments, four pre-trained candidate policies were available, and the zero-shot performance of policy replication inherently depended on the policy chosen as the source. This variation serves as a key motivation for warm-start retraining. While warm-start retraining effectively mitigates efficiency fluctuations caused by source policy selection, the minimal cost of policy parameter replication opens up further opportunities. Specifically, rather than relying solely on single-policy replication, evaluating and selecting the most suitable policy---or dynamically assigning different policies based on specific DC types---is expected to yield even greater performance enhancements.
\par

\section{\uppercase{Conclusion}}\label{sec:translation-conclusion} 
In this study, we proposed sfDA8-X: a strategy-following multi-agent deep reinforcement learning framework that resolves the team-size dependency problem inherent in multi-agent object collection tasks. By using DC, ADC, and three-channel observation matrices that are independent of team size, sfDA8-X allows trained policies to be directly reused across teams of different sizes. This significantly reduces the computational cost of learning compared to previous frameworks, such as sfDA7-X. This architecture enables zero-shot policy transfer to different team sizes via policy replication and facilitates highly efficient warm-start retraining.
\par

Our experimental evaluations demonstrated that zero-shot policy replication maintains high-quality cooperative performance and strategic adherence across varying team sizes without requiring additional training from scratch. Furthermore, when fine-tuned via warm-start retraining, sfDA8-X achieved superior task performance while reducing required training episodes by up to $90\%$. Qualitative heatmap analyses confirmed that the shared policy generated complementary, asymmetric spatial coverage and robust collision avoidance across agents due to input heterogeneity.
\par

Our future work includes developing adaptive source policy selection mechanisms suitable for assigned DCs to further improve and stabilize team performance. We also plan to evaluate the scalability of sfDA8-X in larger teams and wider, more dynamic environments.
\par

\section*{\uppercase{Acknowledgements}}
\noindent\textbf{AI-use disclosure:} OpenAI Codex assisted with Japanese-to-English translation, drafting and revision, evidence organization, and LaTeX checks. The authors verified all technical content, citations, numerical results, and final wording and retain full responsibility for the manuscript.

\bibliographystyle{apalike}
{\scriptsize\bibliography{references}}

\begin{thebibliography}{}

\bibitem[Bernstein et~al., 2002]{bernstein2002complexity}
Bernstein, D.~S., Givan, R., Immerman, N., and Zilberstein, S. (2002).
\newblock {The Complexity of Decentralized Control of Markov Decision
  Processes}.
\newblock {\em Mathematics of Operations Research}, 27(4):819--840.

\bibitem[Dabney et~al., 2018]{dabney2018implicit}
Dabney, W., Ostrovski, G., Silver, D., and Munos, R. (2018).
\newblock {Implicit Quantile Networks for Distributional Reinforcement
  Learning}.
\newblock In {\em {Proceedings of the 35th International Conference on Machine
  Learning}}, pages 1096--1105. PMLR.

\bibitem[Ellis et~al., 2023]{SMAC2}
Ellis, B., Cook, J., Moalla, S., Samvelyan, M., Sun, M., Mahajan, A., Foerster,
  J.~N., and Whiteson, S. (2023).
\newblock {SMACv2: An Improved Benchmark for Cooperative Multi-Agent
  Reinforcement Learning}.
\newblock In {\em Advances in Neural Information Processing Systems (NeurIPS)},
  volume~36, pages 60020--60032.

\bibitem[Fried et~al., 2018]{fried2018speaker}
Fried, D., Hu, R., Cirik, V., Rohrbach, A., Andreas, J., Morency, L.-P.,
  Berg-Kirkpatrick, T., Saenko, K., Klein, D., and Darrell, T. (2018).
\newblock {Speaker-Follower Models for Vision-and-Language Navigation}.
\newblock In {\em Advances in Neural Information Processing Systems},
  volume~31.

\bibitem[Guo et~al., 2025]{guo2025heterogeneous}
Guo, X., Shi, D., Yu, J., and Fan, W. (2025).
\newblock {Heterogeneous Multi-Agent Reinforcement Learning for Zero-Shot
  Scalable Collaboration}.
\newblock {\em Neurocomputing}, 648:130716.

\bibitem[Kurach et~al., 2020]{kurach2020google}
Kurach, K., Raichuk, A., Sta{\'n}czyk, P., Zaj{\k{a}}c, M., Bachem, O.,
  Espeholt, L., Riquelme, C., Vincent, D., Michalski, M., Bousquet, O., and
  Gelly, S. (2020).
\newblock {Google Research Football: A Novel Reinforcement Learning
  Environment}.
\newblock In {\em Proceedings of the AAAI Conference on Artificial
  Intelligence}, volume~34, pages 4501--4510.

\bibitem[Motokawa and Sugawara, 2023]{motokawa2023strategy}
Motokawa, Y. and Sugawara, T. (2023).
\newblock {Strategy-Following Multi-Agent Deep Reinforcement Learning through
  External High-Level Instruction}.
\newblock {\em Procedia Computer Science}, 225:2798--2807.

\bibitem[Samvelyan et~al., 2019]{samvelyan2019starcraft}
Samvelyan, M., Rashid, T., de~Witt, C.~S., Farquhar, G., Nardelli, M., Rudner,
  T. G.~J., Hung, C.-M., Torr, P. H.~S., Foerster, J., and Whiteson, S. (2019).
\newblock {The StarCraft Multi-Agent Challenge}.
\newblock In {\em Proceedings of the 18th International Conference on
  Autonomous Agents and Multiagent Systems (AAMAS)}, pages 2186--2188.

\bibitem[Shalev-Shwartz et~al., 2016]{shalev2016safe}
Shalev-Shwartz, S., Shammah, S., and Shashua, A. (2016).
\newblock {Safe, Multi-Agent, Reinforcement Learning for Autonomous Driving}.
\newblock {\em arXiv preprint arXiv:1610.03295}.

\bibitem[Shi et~al., 2024]{shi2024task}
Shi, H., Tian, Y., Li, H., Huang, J., Shi, L., and Zhou, Y. (2024).
\newblock {Task Offloading and Trajectory Scheduling for UAV-Enabled MEC
  Networks: An MADRL Algorithm with Prioritized Experience Replay}.
\newblock {\em Ad Hoc Networks}, 154:103371.

\bibitem[Sun et~al., 2020]{sun2020program}
Sun, S.-H., Wu, T.-L., and Lim, J.~J. (2020).
\newblock {Program Guided Agent}.
\newblock In {\em International Conference on Learning Representations}.

\bibitem[Takahagi et~al., 2025]{takahagi2025strategy}
Takahagi, Y., Nakasone, G., Motokawa, Y., and Sugawara, T. (2025).
\newblock {Strategy-Following Multi-Agent Deep Reinforcement Learning
  Considering Control Strategies toward Other Agents}.
\newblock In {\em 2025 IEEE/WIC International Conference on Web Intelligence
  and Intelligent Agent Technology (WI-IAT)}, pages 351--358. IEEE.
\newblock DOI: 10.1109/WI-IAT67162.2025.00053.

\bibitem[Vaswani et~al., 2017]{vaswani2017attention}
Vaswani, A., Shazeer, N., Parmar, N., Uszkoreit, J., Jones, L., Gomez, A.~N.,
  Kaiser, L., and Polosukhin, I. (2017).
\newblock {Attention Is All You Need}.
\newblock In {\em Advances in Neural Information Processing Systems},
  volume~30, pages 5998--6008.

\bibitem[Wang et~al., 2024]{wang2024nahteamwork}
Wang, C., Rahman, A., Durugkar, I., Liebman, E., and Stone, P. (2024).
\newblock {$N$-Agent Ad Hoc Teamwork}.
\newblock In {\em Advances in Neural Information Processing Systems},
  volume~37.

\bibitem[Wang et~al., 2023]{ye2023mutual}
Wang, J., Ye, D., and Lu, Z. (2023).
\newblock {Mutual-Information Regularized Multi-Agent Policy Iteration}.
\newblock In {\em Advances in Neural Information Processing Systems},
  volume~36.

\bibitem[Wu et~al., 2023]{wu2023toward}
Wu, J., Huang, Z., Hu, Z., and Lv, C. (2023).
\newblock {Toward Human-in-the-Loop AI: Enhancing Deep Reinforcement Learning
  via Real-Time Human Guidance for Autonomous Driving}.
\newblock {\em Engineering}, 21:75--91.

\end{thebibliography}

\end{document}